# Deformation mechanism map of Cu/Nb nanoscale metallic multilayers as a function of temperature and layer thickness

J. Snel[1], M. A. Monclús[1], M. Castillo-Rodríguez[1], N. Mara[2],
I. J. Beyerlein[3], J. LLorca[1,4] and J. M. Molina-Aldareguía[1]

[1] IMDEA Materials Institute, c/Eric Kandel 2, 28906 Getafe, Madrid, Spain
[2] Institute for Materials Science and Center for Integrated Nanotechnologies, Los Alamos National Laboratory, Los Alamos, NM 87545, USA
[3] University of California, Santa Barbara, CA 93106-5070
[4] Department of Materials Science, Polytechnic University of Madrid. E.T.S. de Ingenieros de Caminos, 28040 Madrid, Spain

**Abstract**

The mechanical properties and deformation mechanisms of Cu/Nb nanoscale metallic multilayers (NMMs) manufactured by accumulative roll bonding (ARB) are studied at 25 ºC and 400ºC. Cu/Nb NMMs with individual layer thicknesses between 7 and 63 nm were tested by *in-situ* micropillar compression inside a scanning electron microscope Yield strength, strain-rate sensitivities and activation volumes were obtained from the pillar compression tests. The deformed micropillars were examined under scanning and transmission electron microscopy in order to examine the deformation mechanisms active for different layer thicknesses and temperatures. The analysis suggests that room temperature deformation was determined by dislocation glide at larger layer thicknesses and interface-related mechanisms at the thinner layer thicknesses. The high temperature compression tests, in contrast, revealed superior thermo-mechanical stability and strength retention for the NMMs with larger layer thicknesses with deformation controlled by dislocation glide. A remarkable transition in deformation mechanism occurred as the layer thickness decreased, to a deformation response controlled by diffusion processes along the interfaces, which resulted in temperature-induced softening. A deformation mechanism map, in terms of layer thickness and temperature, is proposed from the results obtained in this investigation.

## 1. Introduction

Nanoscale metallic multilayers (NMMs) formed by alternating layers of two dissimilar metals with layer thicknesses in the nanoscale range have very attractive properties including high hardness, strength and thermal stability. The outstanding mechanical performance is mainly controlled by the high density of interfaces that can act as dislocation sources, sinks, storage sites and promote or demote dislocation transmission to adjacent layers [1]. When the individual layer thicknesses decrease to nanoscale sizes, the interfaces dominate the dislocation-based deformation processes.

Amongst them, Cu/Nb NMMs are one of the most widely studied systems, not only for their high strength [2–4] and thermal stability [5,6], but also for their high radiation tolerance [7]. This latter behaviour is attributed to the capacity of the Cu/Nb interfaces to operate as sinks for radiation-induced defects, so the strength of the material is maintained after irradiation. Its potential applications (i.e. increased energy efficiency in nuclear environments) often require good mechanical properties at elevated temperatures. However, the information available on the mechanical behaviour of Cu/Nb NMMs has been focussed on room temperature behaviour, while high temperature studies are very limited and related to the tensile properties [8–10] and the thermal stability [5,11,12].

Cu/Nb NMMs can be fabricated by accumulative roll bonding (ARB) and physical vapour deposition (PVD), leading to NMMs with different Cu/Nb interface structure. The features of Cu/Nb interfaces and their interaction with dislocations dictate their mechanical properties and the mechanisms that lead to plastic instabilities and slip/shear band formation at room temperature have been studied in detail [13]. In the particular case of ARB Cu/Nb NMMs, nanoindentation tests at high temperature (up to 400 °C) revealed that the optimum strength is achieved at a critical layer thickness that shifted from ≈ 10 nm at room temperature to ≈ 18 nm at 400°C. This layer thickness marks the transition from thermally activated glide to dislocation transmission across the layers at each temperature [14]. However, the difficulties associated with the analysis and interpretation of the nanoindentation tests did not allow obtaining a comprehensive map of the deformation mechanisms that control the mechanical behaviour of these NMMs as a function of layer thickness and temperature and this is the main objective of this investigation.

## 2. Material and methods

### 2.1. Material

Cu/Nb NMMs were prepared by ARB, a process that involves repeated rolling, sectioning, stacking, bonding, and re-rolling [15]. The Cu/Nb stack consists initially of alternating Cu (99.99% purity) and Nb (99.94% purity) layers of equal thickness. Four Cu/Nb NMM sample foils were produced with average individual layer thickness ($h$) of ≈ 7, 16, 34 and 63 nm. In order to ensure that the ARB microstructures were free of any surface effects from the final synthesis step, samples were cut out from the foil, mechanically ground (to about 1/3 of the total thickness from one side) with 1200 and 2000 grit size SiC paper, and finally polished with diamond paste (1 and 0.3 μm particle size) to prepare the surface prior to FIB milling The final average roughness ($R_a$) was ≈ 10 nm.

2.2. Micropillar compression

Cu/Nb micropillars, with a square cross-section, were machined by focused ion beam (FIB) milling using a FEI Helios Nanolab 600i dual beam FIB-FEGSEM operated at 30 kV. In the first step, a large trench was machined at currents of 47 nA to provide sufficient clearance for the flat-punch indenter and to allow viewing of the micropillar along its full length. Milling continued in the second and third steps at currents of 9.7 nA and 2.5 nA, respectively. Finally, the pillar taper was reduced at currents of 0.79 nA. The dimensions of the final micropillars were in between 4.7 and 5.6 μm in lateral size, with aspect ratios in the range 2.5-2.9 and a taper of less than 1.0°.

The room temperature compression experiments were performed with a Hysitron TI950 TriboIndenter$^{TM}$ and a 10 μm flat punch diamond indenter. The high temperature experiments were performed with a Hysitron PI87HT SEM picoindenter installed inside a Zeiss EVO MA15 SEM. Room temperature compression results for both systems were compared for reproducibility and were found satisfactory. Constant strain-rate and strain-rate jump tests were performed at displacement rates of 1.5, 15 and 150 nm/s corresponding to strain-rates of $10^{-4}$, $10^{-3}$ and $10^{-2}$ s$^{-1}$. In strain rate jump (SRJ) tests, the change in stress resulting from a sudden jump in the imposed strain rate is used to determine the strain rate sensitivity and activation volumes. Up to five strain jumps were introduced in each compression test, starting from an initial strain rate of $10^{-3}$ s$^{-1}$. The first jump was introduced at an applied strain of 5% and, afterwards, at steps of 2.5% applied strain in the following order: $10^{-3}$ s$^{-1}$ → $10^{-4}$ s$^{-1}$ → $10^{-3}$ s$^{-1}$ → $10^{-2}$ s$^{-1}$ → $10^{-3}$ s$^{-1}$. The main advantages of the SRJ method over the constant strain rate (CSR) method are twofold. Firstly, the strain rate sensitivity can be determined from a single compression experiment by introducing several strain jumps, as shown in Fig. 1, and this strategy minimizes the inherent scatter associated with nanomechanical testing. Secondly, the reduction in testing time eliminates problems associated with thermal drift [16].

The strain rate sensitivity, *m*, can be calculated from the flow stress and the strain rate before and after each jump, as depicted in the inset of Fig. 1, according to

$$m = \frac{ln(\sigma_2/\sigma_1)}{ln(\dot{\varepsilon}_2/\dot{\varepsilon}_1)}\bigg|_T \qquad (1)$$

The activation volume, *V*, is given by

$$V = \sqrt{3}kT\frac{\partial \ln(\dot{\varepsilon})}{\partial \sigma_f} \qquad (2)$$

where $k$ is the Boltzman constant, $T$ the absolute temperature and $\sigma_f$ the flow stress. The specific activation volume, $V^*$, is given by

$$V^* = \frac{V}{b^3} \qquad (3)$$

where $b$ is the value of the Burgers vector.

Sneddon´s correction for instrument and substrate compliance was applied to the load-displacement curves [17]. Stress-strain curves were obtained from the corrected load-displacement curves assuming that the pillars were perfect prisms of length $L_0$ and square cross-sectional area $A_0$ measured as the top cross-sectional area of the pillar. The engineering stress $\sigma_{eng}$ and engineering strain $\varepsilon_{eng}$ were calculated, respectively, as

$$\sigma_{eng} = \frac{P}{A_0} \qquad (4)$$

$$\varepsilon_{eng} = \frac{u_p}{L_0} \qquad (5)$$

where $P$ is the applied force and $u_p$ the corrected pillar displacement.

Post-compression images of the pillars were obtained with the FEI Helios Nanolab 600i dual beam FIB-FEGSEM. TEM samples of the compressed micropillars were also prepared with the same FIB instrument. Post-compression TEM images and energy dispersive X-Ray spectroscopy (EDS) analysis of the deformed pillars were taken with an FEI Talos F200X TEM.

### 3. Experimental results

3.1. Room temperature micropillar compression

Representative stress-strain curves from CSR and SRJ tests at room temperature are plotted in Fig. 1 as a function of layer thickness. The strength of the Cu/Nb MNMs increased as the layer thickness ($h$) decreased. The highest strengths (over 2 GPa) were achieved for $h = 7$ and 16 nm and they dropped to approximately 1.7 and 1.3 GPa for $h = 34$ and 63 nm, respectively. The finer layer thickness NMMs ($h = 7$ and 16 nm) were more prone to plastic instabilities as indicated by the onset of softening in some of the stress-strain curves. This behaviour is expected from the limited strain-hardening capability of these materials that might lead to strain localization.

Representative SEM images of pillars compressed up to an applied strain between 10 and 15% are shown in Fig. 2. Different deformation modes are observed as a function of layer thickness. The pillar with $h = 63$ nm deformed up to 15 % strain displays some bulging (Fig. 2d) which is more apparent at mid height along the length of the pillar. This change indicates that any possible enhancement in applied stresses at the upper part of the pillar due to the taper is negligible, and that friction with the flat punch introduces some constraint on the deformation of this part of the pillar. The layers in the bulged region appear to co-deform homogeneously parallel to the interfaces, without the formation of any sharp shear bands that might break across the layers. This behaviour is significantly different from the one observed for $h$ = of 7 nm and 16 nm, where shear failure takes place at smaller strains of ≈ 10%. Shear failure was observed to occur along one or several shear planes, initiating in one of the upper corners and propagating downwards with angles in the range of 32-38º. Interestingly, the shear plane was always inclined towards the rolling direction, indicating a relation between the preferred shearing direction and foil texture. In the case of Cu/Nb with $h$ = 34 nm, a transition in the deformation behaviour was observed: from homogeneous deformation occurring up to strains ≈ 10 % to the onset of shear failure observed at strains of ≈ 15 %. A shear band developed at this strain, as indicated by the white arrow in Fig. 2c. Comparing the stress-strain curves with these images, shear failure was linked to the higher yield stresses and lower strain hardening capacity found in the thinner layers, and it is probably responsible for the softening observed in the stress-strain curves. The more homogeneous deformation observed in the 63 nm Cu/Nb pillars was linked with lower flow stress and higher strain hardening capacity.

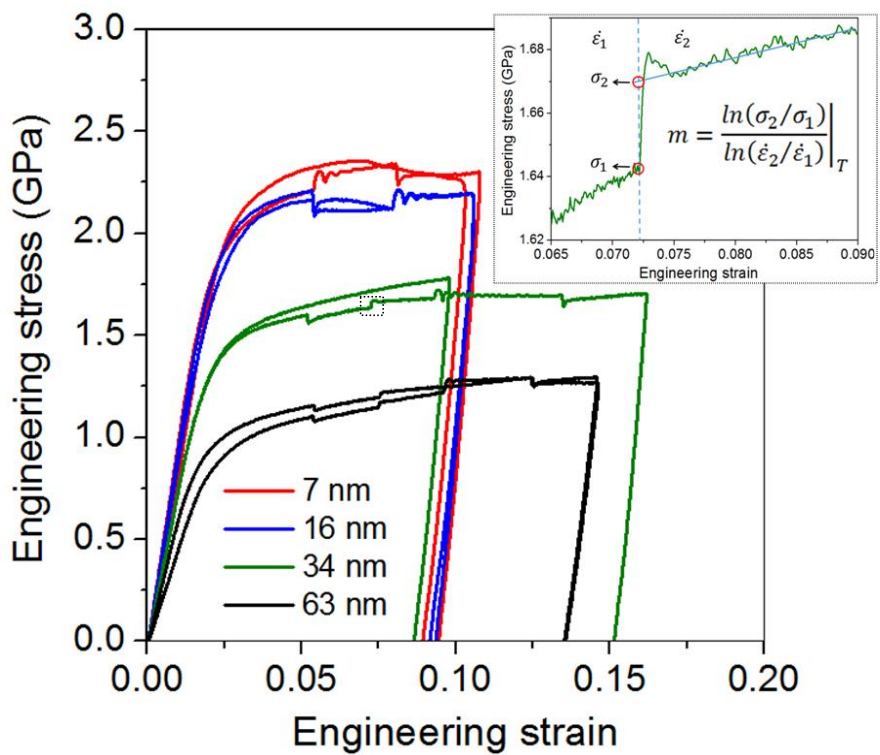

**Fig. 1.** Representative engineering stress-strain curves obtained from compression of ARB Cu/Nb micropillars at room temperature with average layer thickness $h$ = 7, 16, 34 and 63 nm. The inset shows the method for extracting the strain-rate sensitivity from the SRJ tests: a linear curve is fitted to the constant strain-rate segment after the jump and extrapolated to the jump strain. Stress levels $\sigma_1$ and $\sigma_2$ are measured and the strain-rate sensitivity is calculated with Equation (1).

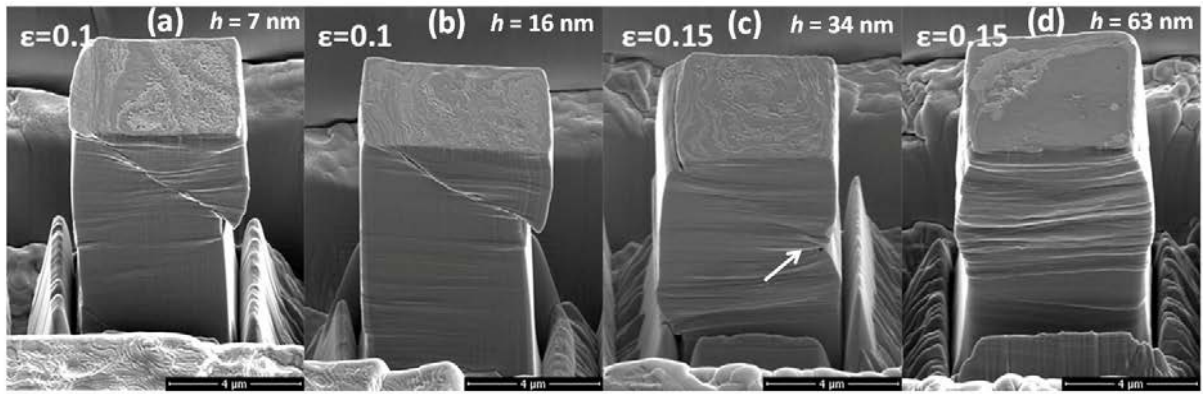

**Fig. 2.** SEM images of Cu/Nb micropillars with different layer thicknesses, $h$, deformed at room temperature: (a)-(b) $h$ = 7 and 16 nm pillars compressed to strains of 0.1 and (c)-(d) $h$ = 34 and 63 nm pillars compressed to strains of 0.15. The white arrow indicates the presence of a shear band.

Bright-field (BF) TEM images of the Cu/Nb NMMs with $h$ = 34 nm deformed up to $\varepsilon$ = 0.15 at room temperature are shown in Figs 3a-b. Fig. 3a confirms the nucleation of a shear band at the upper left corner of the micropillar. The shear band propagates downwards and accumulates shear through the local rotation of the layers, but without breaking through them, and the layers remain continuous across the diffuse boundaries of the shear band (Fig. 3b). In addition, two cracks nucleated at the intersections of the shear band with the edges of the pillar are shown in Fig. 3a. They will eventually propagate and induce shear failure of the pillar. The mechanisms of shear band formation and final shear failure are analogous in the case of the 7 and 16 nm thick Cu/Nb NMMs, but they take place at larger applied strains than for the 34 nm thick NMM. A high volume fraction of deformation twins within the copper layers close to the shear band is observed in Fig. 3c. The width of the twins is in the range 5-10 nm, consistent with the widths reported by Zheng et al. [18].

3.2. High temperature micropillar compression

Representative stress-strain curves obtained from strain rate jump (SRJ) compression tests at 25°C and 400°C using the *in-situ* device are compared in Fig. 4 as a function of layer thickness. The NMMs with thinner layers in Fig. 4a and b display a larger drop in strength with temperature than the NMMs with larger layer thicknesses in Fig. 4c and d, in agreement with previous hot hardness reports [14]. The size of the strain rate jumps also suggests a larger increase in the strain rate sensitivity (SRS) with temperature for the NMMs with thinnest layers.

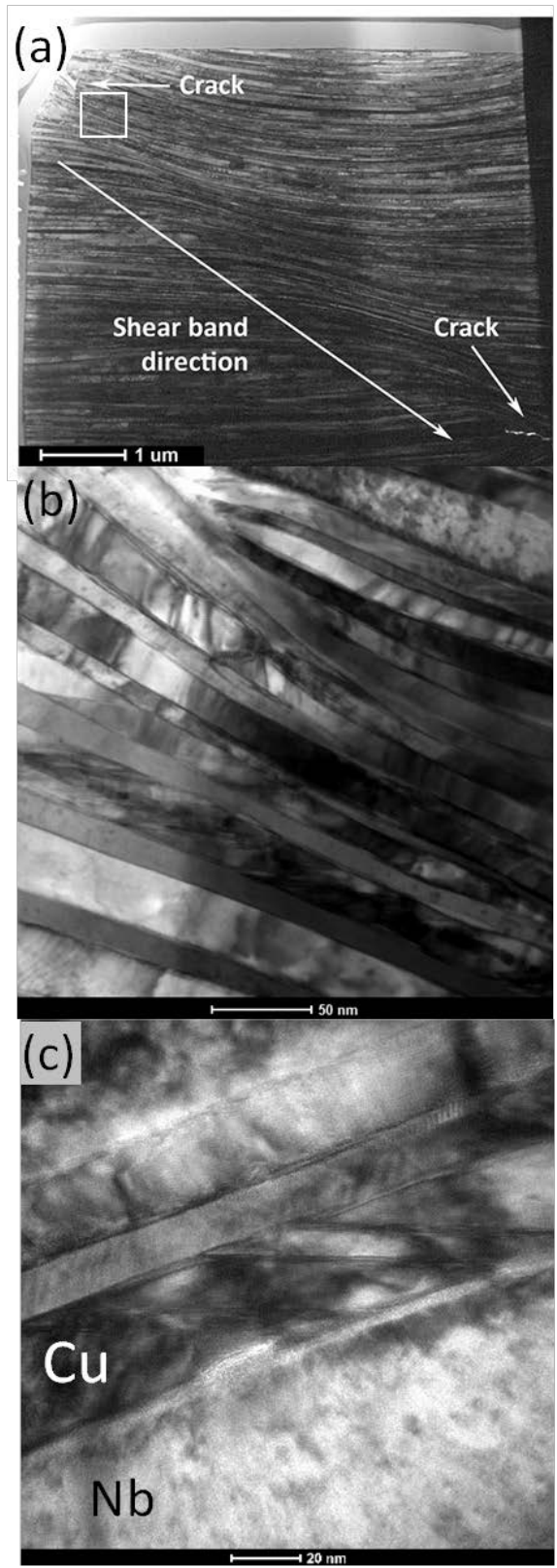

**Fig. 3.** (a) Bright –field TEM image of a cross- section of the 34 nm ARB Cu/Nb pillar compressed at room temperature showing a shear band across the sample and two cracks at each end of the band. (b) High magnification TEM image of the area delimited by the square in Fig. 3a showing that Cu and Nb layers were continuous across the shear band. (c) Detailed view of nanotwins formed in the Cu layer.

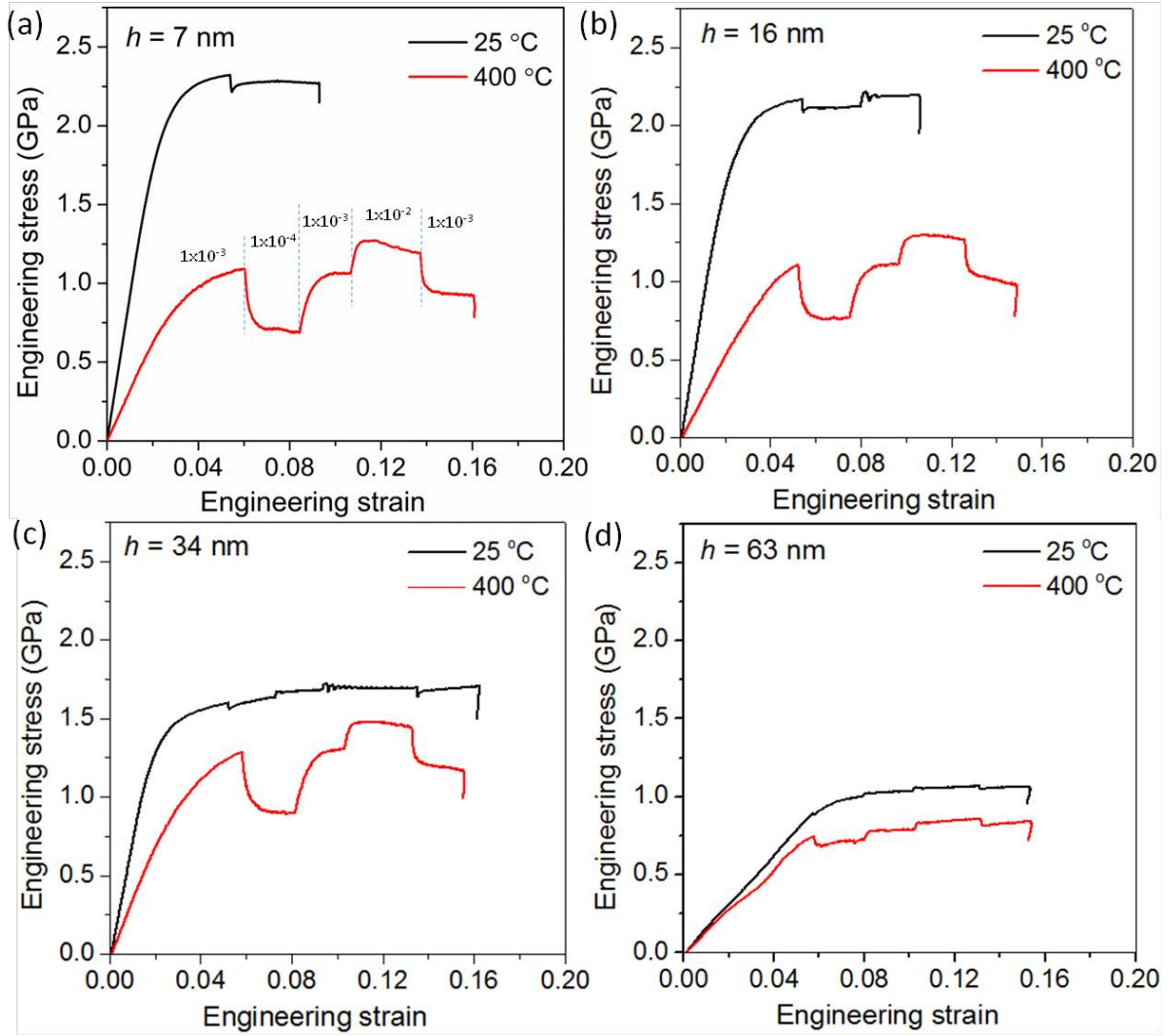

**Fig. 4.** Representative stress-strain curves of Cu/Nb micropillars with average layer thickness $h = 7$ nm (a), 16 nm (b), 34 nm (c) and 63 nm (d) obtained from SRJ compression tests at 25 °C and 400 °C. Strain-rates were applied in the following order: $10^{-3}$ s$^{-1}$ → $10^{-4}$ s$^{-1}$ → $10^{-3}$ s$^{-1}$ → $10^{-2}$ s$^{-1}$ → $10^{-3}$ s$^{-1}$, as indicated in (a).

SEM images of Cu/Nb micropillars with different layer thicknesses compressed at 25 °C and 400 °C are shown in Fig. 5. The ones corresponding to 63 nm thick Cu/Nb micropillars compressed at room temperature in the stand-alone instrument and the in-situ device are depicted in Fig. 2d and Fig. 5g, respectively. It appears that the latter suffered a small misalignment problem, as revealed by the tendency of the pillar to bend slightly. The misalignment problem was not however, severe enough to change the stress-strain curve, which was identical in both cases. In addition, the layers in Fig. 5g still appear to co-deform homogeneously, as was concluded from Fig. 2d. Regarding the compression at 400ºC, there are no obvious differences in the deformation morphology with temperature for the 63 nm layer thick Cu/Nb NMMs (compare Fig. 5g at 25ºC with Fig, 5h at 400ºC). These results suggest that the underlying deformation mechanisms do not change between 25ºC and 400ºC in this case.

Changes in the deformation with temperature were dramatic, however, for the NMMs with the thinnest layer thicknesses of 7 and 16 nm. Material can be seen to extrude from the pillar side walls, in the form of droplets or globules at 400ºC (Figs. 5b and 5d). While the number of globules on the side walls was negligible for the 63 nm thick layers in Fig. 5h, it increases considerably as the layer thickness was reduced, especially for the 7 nm thick layers in Fig. 5b. Moreover, their density is larger at the higher stressed areas, like the upper part of the pillar in Fig. 5d. So, the globules seem to develop in response to the applied stress as a result of stress-assisted diffusion processes at 400ºC.

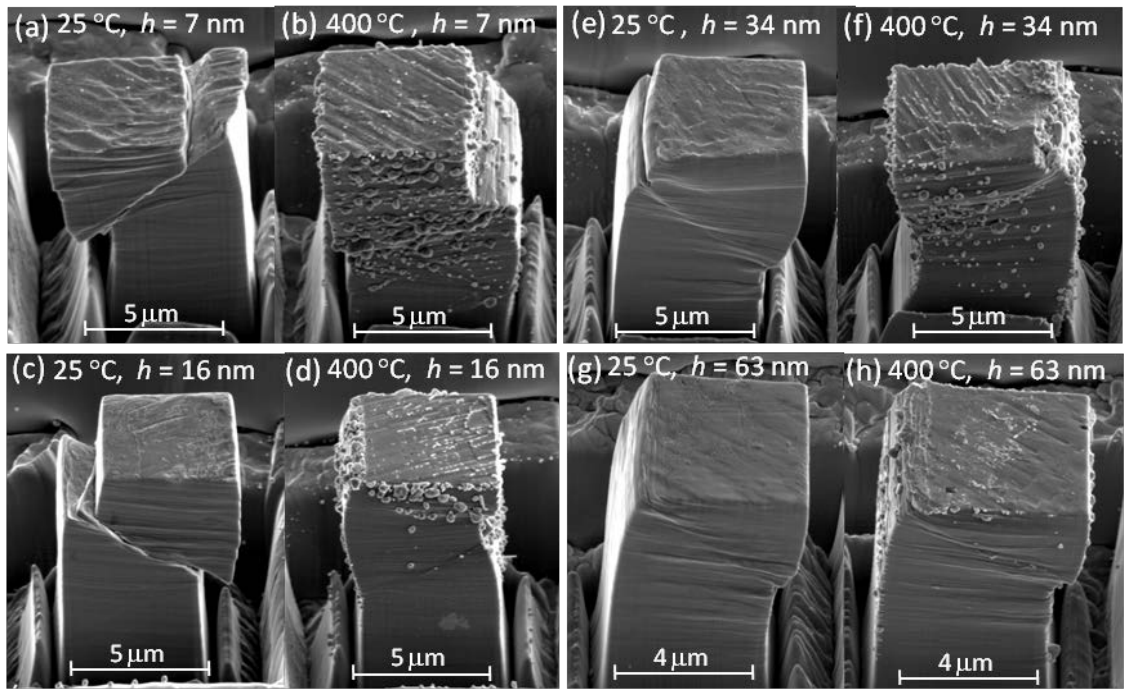

**Fig. 5.** SEM images of deformed micropillars compressed at 25°C and 400°C, with individual layer thicknesses, *h*, of (a-b) 7 nm, (c-d) 16 nm (e-f) 34 nm and (g-h) 63 nm.

The onset of stress-assisted diffusion at 400ºC renders a more homogeneous deformation distribution along the length of the pillar. As a result, shear failure took place in the 7 nm and 16 nm thick NMMs deformed at 25ºC (Figs. 5a and 5c), but it did not happen in the homogenously deformed pillars at 400ºC (Figs. 5b and 5d). Nevertheless, the pillars deformed at 400ºC were also prone to the formation of shear bands. This is depicted in Fig. 6, which shows BF TEM images of the 16 nm layer thick Cu/Nb deformed at 400 °C. A narrow shear band is clearly observed in Fig. 6a, but, there is no evidence of the formation of shear cracks, contrary to the room temperature observations. Moreover, the layers are disrupted across the shear band at 400ºC, another behaviour that was not found at room temperature (Fig. 3a). Higher magnification images of the highly deformed shear band in Fig. 6b and c, which correspond to the rectangles highlighted in Fig. 6a, show clear evidences of re-crystallization and grain growth. They confirm again the active role of diffusional processes on the deformation behaviour.

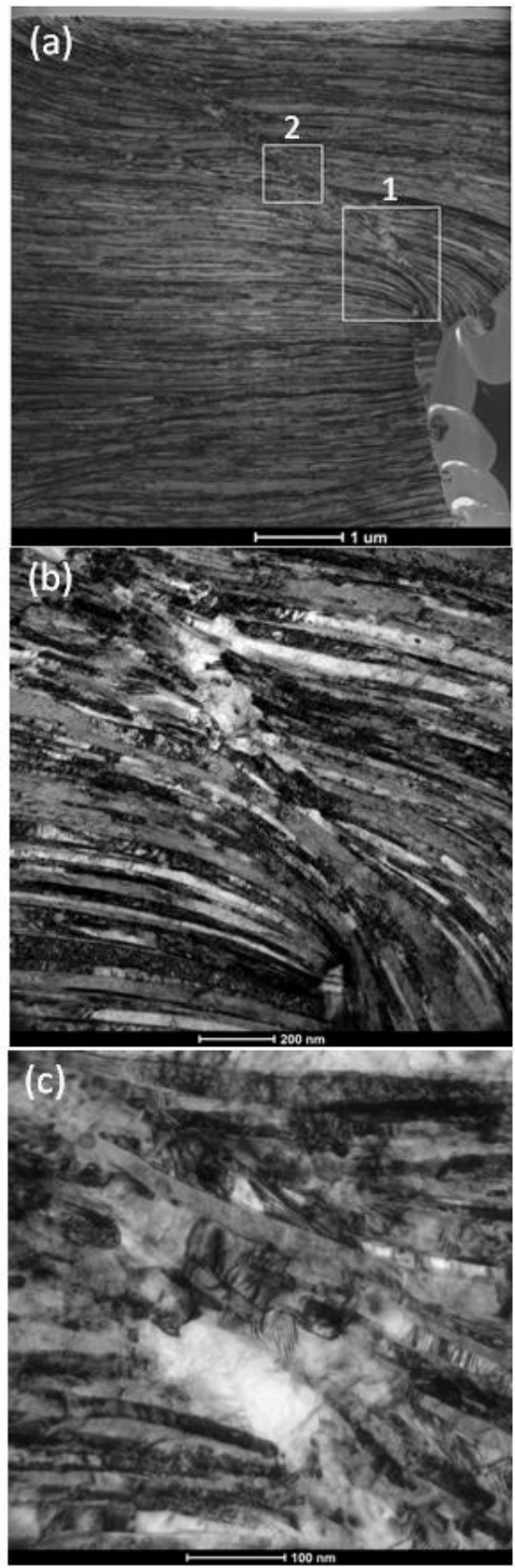

**Fig. 6.** (a) BF-TEM cross-sectional images of Cu/Nb micropillar with $h$ = 16 nm compressed at 400°C. (b) and (c) Higher magnification TEM images of squared areas 1 and 2, respectively, showing layer discontinuity and recrystallization at the shear band.

HAADF-STEM images alongside EDX maps of the 16 nm layer thick Cu/Nb pillar compressed at 400°C are shown in Fig. 7. Fig. 7a corresponds to areas at the side wall, far away from the shear band. The layers deform homogeneously and the globules that extrude out correspond to Cu microcrystals, as determined by the EDX map of Fig. 7b. Within the highly deformed shear band area of Fig. 7c, the Nb layers seem to break up and Cu diffuses to fill up the cracks, as seen in the EDX map of Fig. 7d. This healing process apparently contributes to hinder the formation of shear cracks. Thus, TEM observations indicate that the stress-assisted diffusion processes take place in the softer Cu layers, and not in the Nb layers. Similar processes have also been reported at 100ºC in Al/SiC nanolaminates [19] and at 200ºC in Cu/TiN [20] and Cu/Cr [21] nanolaminates. The fact that diffusional process take place preferentially in the Cu layers and not in the Nb layers is not surprising because 400ºC represents a homologous temperature of 0.14 for Nb and 0.3 for Cu.

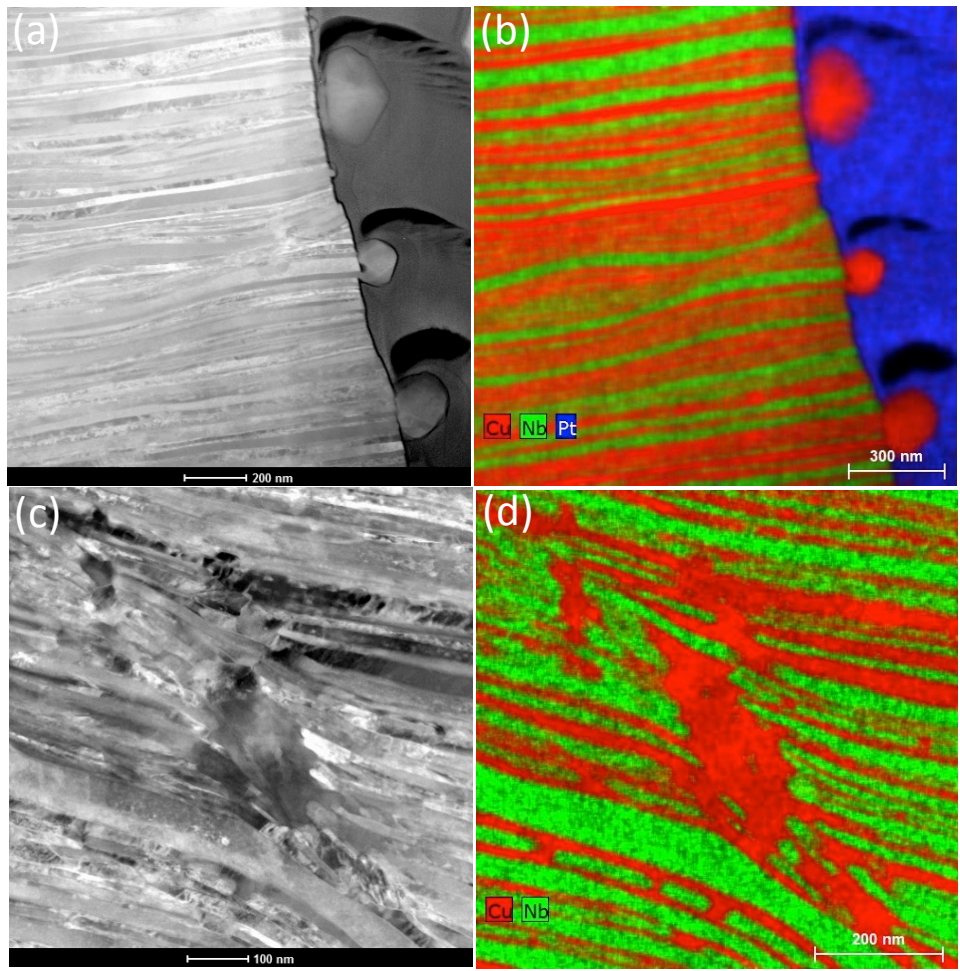

**Fig. 7.** (a) HAADF-STEM image of the 16 nm Cu/Nb pillar compressed at 400 °C showing homogeneous deformation of the layers. (b) EDX mapping image showing the formation of Cu microcrystals at the pillar side walls. (c) HAADF-STEM image of the highly deformed area with crystal growth disrupting the continuity of the layers. (d) EDX map of (c) showing fracture of the Nb layer and Cu filling the cavity formed during deformation.

## 4. Discussion

4.1. Yield stress variation with layer thickness as a function of temperature

The yield stress, $\sigma_{0.2\%}$ (measured at strain rates of $10^{-3}$ s$^{-1}$ at 0.2 % strain offset from the engineering stress-strain curves) of the Cu/Nb NMMs is plotted in Fig. 8 as a function of layer thickness and temperature. As expected, the yield stress increases at room temperature as the layer thickness decreases. A decrease in yield stress was observed at 400ºC for all layer thicknesses, yet the NMMs retain a remarkable yield stress of ≈ 780 MPa and ≈ 600 MPa for $h$ ≈ 16 and 63 nm. These values are in qualitative agreement with separate studies that report a yield stress of ≈ 700 MPa in tension at 500°C for PVD Cu/Nb NMMs with $h$ ≈ 60 nm [8]. Intriguingly, a relatively larger drop in strength is observed for the thinnest multilayer with $h$ = 7 nm, indicating a possible change of deformation mechanism. As a result, the peak in yield stress shifts from 1.74 GPa for $h$ = 7 nm at room-temperature to 0.9 GPa for $h$ = 16 nm at 400ºC, in agreement with earlier hot hardness observations [14].

It is widely accepted that Cu/Nb NMMs with layer thicknesses between a few tens to a few hundred of nanometres deform by confined layer slip (CLS) at room temperature [2]. Deformation in CLS is controlled by the glide of single dislocation loops in the Cu layers that are pinned at the Cu/Nb interfaces. The stress, $\sigma_{CLS}$, necessary to propagate a glide loop of Burgers vector $b$ confined to the softer Cu layer is given by:

$$\sigma_{CLS} = M \frac{\mu b}{8\pi h'} \left(\frac{4-\nu}{1-\nu}\right) \left[\ln \frac{\alpha h'}{b}\right] - \frac{f}{h} + \frac{C}{\lambda} \qquad (6)$$

where $M$ is the Taylor factor, $\mu$ the shear modulus of Cu, $h'$ (=$h$/sin$\phi$) the thickness of the Cu layer parallel to the glide plane ($\phi$), $\nu$ the Poisson's ratio of Cu, $\alpha$ the core cut-off parameter, $f$ the characteristic interface stress of the multilayer, which has a typical value of 2 J/m$^2$ in NMMs with incoherent interfaces, and $C$ the stress field of a single array of edge dislocations spaced by $\lambda$ given by $C = \mu b/(1-\nu)$ where $\lambda$ is related to the applied (yield) strain $\varepsilon$ by $\lambda = b/\varepsilon$. The continuous black line in Fig. 8 corresponds to the CLS model yield stress prediction ($\varepsilon$ = 0.2%) for $M$ = 3.1, $\mu$ = 48 GPa, $\nu$ = 0.3, $b$ = 0.25 nm, $\phi$ = 70.5° and a core cut-off parameter $\alpha$ = 1. The prediction is in remarkably good agreement with the experimental results for the 63 and 34 nm thick Cu/Nb NMMs, but the experimental yield stress reaches a plateau for $h$ < 34 nm that is considerably lower than the predictions of the CLS model. It is widely accepted that this plateau is reached for very small layer thicknesses when the CLS stress ($\sigma_{CLS}$) exceeds the stress required for dislocation transmission (DT) across the interfaces ($\sigma_T$). According to Fig. 8, $\sigma_T$ can be estimated to be around ≈ 1.7 GPa in this case. The transition from CLS to DT occurs at layer thicknesses of a few nm for PVD Cu/Nb [22], but this transition has been observed to occur at larger thicknesses ( > 10 nm) for ARB Cu/Nb [14], in agreement with the current results. This transition is compatible with the deformation behaviour observed in the micropillars. The relatively homogeneous deformation observed for *h* = 63 nm pillars agrees with a CLS-type mechanism, while the shear failure observed for the 7 and 16 nm thick micropillars is indicative of the activation of dislocation transmission across interfaces, as suggested by Misra et al. [23]. They proposed that the shear localization resulted from the passage of several dislocations across the interface at the site of slip. Moreover, the transition from uniform deformation to shear localization observed for the 34 nm thick multilayer at higher applied strain also agrees with this view. In this case, the NMM initially yields at a stress of ≈ 1.2 GPa presumably by a CLS mechanism, but the maximum flow stress reaches ≈ 1.7 GPa for an applied strain of 10% due to strain hardening, which is close to the stress required for dislocation transmission at interfaces.

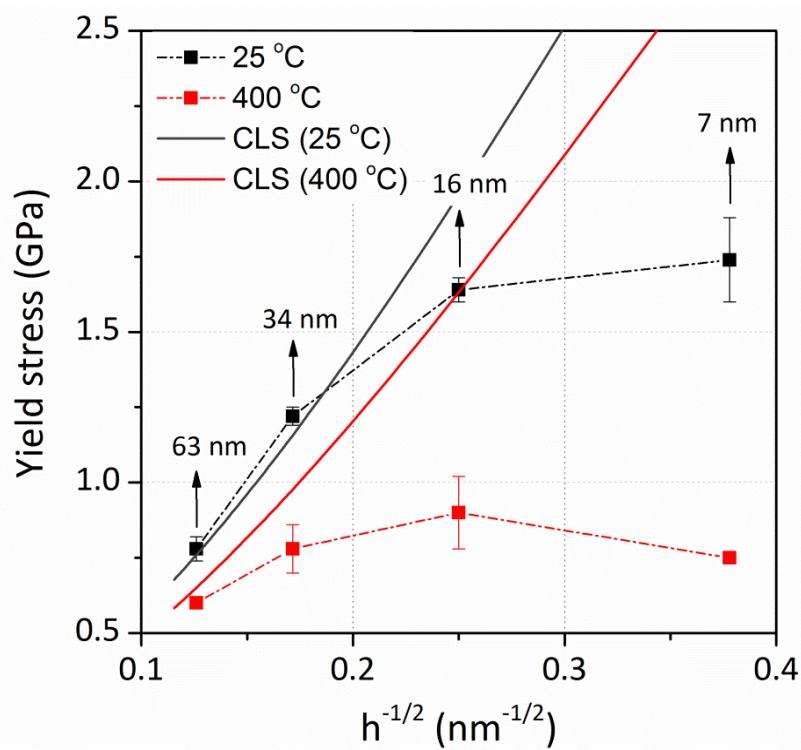

**Fig. 8.** Yield strength as a function of $h^{-1/2}$ for temperatures of 25 °C and 400 °C.

Regarding the high temperature results, the red line in Fig. 8 shows the predictions of the CLS model with the Cu shear modulus at 400ºC (= 40 GPa [24]). Only the experimental result for the thick 63 nm multilayer agrees well with the predicted yield stress at this temperature. This result, together with the identical morphologies of the pillars deformed at 25ºC (Fig. 5g) and 400ºC (Fig. 5h), indicates that the same CLS deformation mechanism prevails at 400ºC for the NMM with 63 nm layer thickness. The thinnest NMMs are, however, much softer than expected from the CLS model and, as discussed before, the reduction in strength increases as the layer thickness decreases. Several factors can be responsible for this behaviour, such as microstructural changes leading to a breakdown and coarsening of the multilayer structure or the onset of thermally-activated deformation mechanisms. Microstructural changes with temperature can be quickly discarded based on the SEM and TEM observations of the pillars deformed at 400ºC. As a matter of fact, ARB Cu/Nb NMMs with $h$ = 18 nm have been shown to be remarkably thermally stable after 1 h heat treatment at temperatures up to 700 ºC [5]. Therefore, the results suggest the deformation mechanisms for the thinnest multilayers at 400ºC are controlled by other mechanisms. Several thermally-activated processes can be considered, such as thermally-activated dislocation transmission across the interfaces, dislocation climb or diffusion. The formation of Cu globlues along the side length of the pillars points to stress-assisted diffusion mechanisms. Several observations suggest that the amount of diffusional processes increases dramatically as the layer thickness decreases. Firstly, the number of globules extruded out at the side walls was negligible for the 63 nm thick layers in Fig. 5h, and increased considerably for smaller layer thicknesses, especially for 7 nm thick layers in Fig. 5b. Secondly, the yield stress shows an inverse size effect by which the 7 nm thick multilayer is softer than the 16 nm thick one for the same applied strain. These observations point to grain boundary/interface diffusion, rather than lattice diffusion, as the controlling diffusion process, following a mechanism that can be either similar to Coble creep [25], dislocation climb or grain boundary sliding [26]. The Cu layers contain only one grain through the thickness in the ARB Cu/Nb NMMs and the lateral dimensions of the grains in the rolling and transverse directions are at least one order of magnitude larger than the layer thickness [15]. Thus, it is reasonable to assume that the Cu/Nb interfaces are the preferred diffusion paths. Assuming the same diffusion coefficient as that for grain boundary diffusion in Cu at 400ºC, $D_{gb}$ (400ºC)=4 x $10^{-13}$ m$^2$/s [27], the diffusion distance over the typical duration of a compression test ($\approx$ 100 s) amounts to $x \sim \sqrt{D_{gb}t} \sim$ 6.3 μm. This distance is of the same order of magnitude than the pillar dimensions, rendering stress-assisted grain-boundary diffusion at the testing conditions likely. In comparison, the lattice diffusion coefficient of Cu at 400ºC, $D_{lat}$ (400ºC)=2 x $10^{-16}$ m$^2$/s [28], for which the diffusion distance $x \sim$ 0.14 μm. The strain rate sensitivity can shed more light into the operating deformation mechanisms in each regime.

4.2. Strain-rate sensitivity and activation volume

Strain-rate sensitivities (SRS) and activation volumes ($V$) for Cu/Nb as a function of layer thickness and temperature are plotted in Fig. 9. Values were determined for the $10^{-4} - 10^{-3}$ s$^{-1}$ (Figs. 9a and 9c) and for the $10^{-3} - 10^{-2}$ s$^{-1}$ (Figs. 9b and 9d) strain-rate jumps, which from here on are referred to as "slow" and "fast" compression rates.

A relatively small SRS was observed for all layer thicknesses at room temperature. SRS values were on the order of 0.01 - 0.06, which are 3 to 4 times lower than those observed in nc-Cu with nanoscale grain sizes [20, 21]. This implies stress exponents $n$ of the order of 100, which is roughly what is expected for coarse-grain fcc metals at high stresses and low temperatures [31], where the rate controlling deformation mechanisms is dislocation glide. Interestingly, a slight increase in SRS is observed in the thinnest multilayers (7 nm and 16 nm) (indicated by blue arrows in Figs. 9a-b). The activation volumes, in the range 17-70 $b^3$, scale with layer thickness, which is compatible with confined layer slip (10–100 $b^3$) [32].

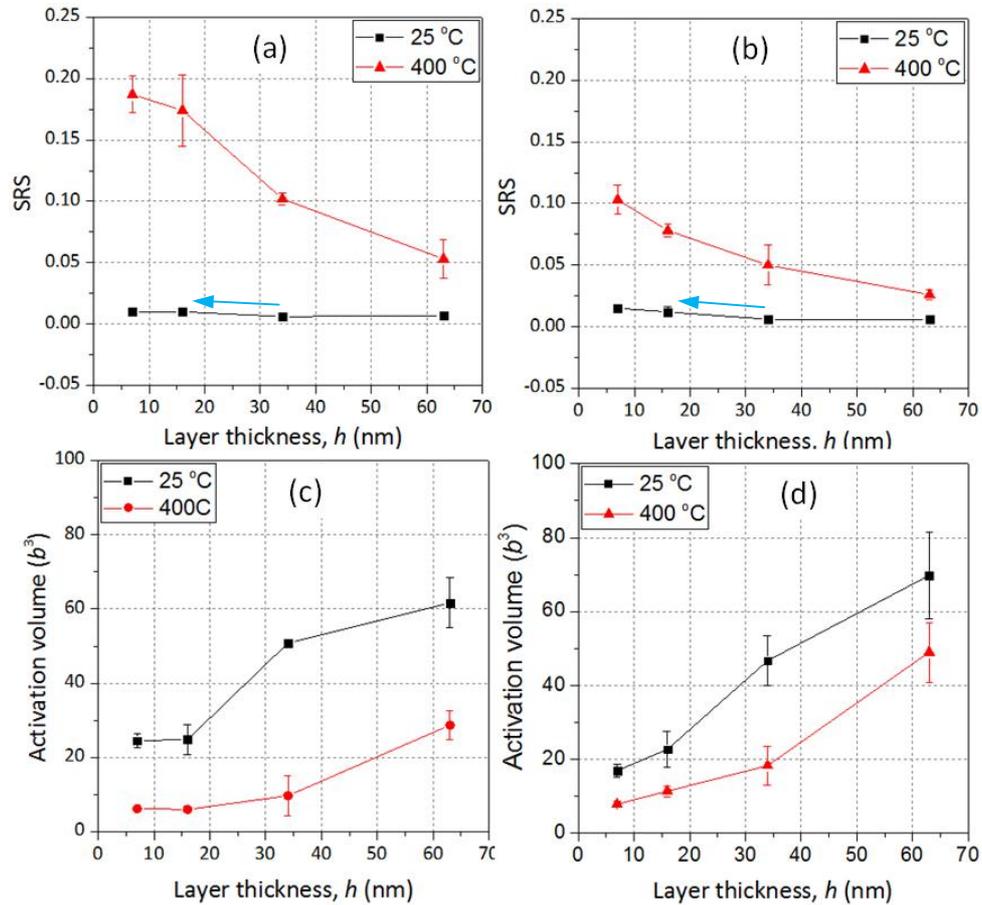

**Fig. 9.** Strain-rate sensitivity (SRS) as a function of layer thickness at 25 °C and 400 °C for (a) $10^{-4}$-$10^{-3}$ s$^{-1}$ and (b) $10^{-3}$-$10^{-2}$ s$^{-1}$ strain-rate regimes. Activation volume, $V$, as a function of layer thickness at 25 °C and 400 °C for (c) $10^{-4}$-$10^{-3}$ s$^{-1}$ and (d) $10^{-3}$-$10^{-2}$ s$^{-1}$ strain-rate regimes.

With increasing temperature, the SRS of the thicker multilayer, with $h$ = 63 nm, increased only gradually to an average value of ≈ 0.04. This behavior suggests that dislocation glide controlled deformation prevails in this case. A large increase in strain-rate sensitivity was observed, however, for the thinner multilayers, $h$ < 34 nm, at 400°C. The SRS values approached 0.2 for the thinnest layer thickness and the slowest compression rates (Figure 9a), and the activation volumes were small (of the order of ≈ 10 $b^3$). These data suggest that deformation is entering a power-law creep regime in this case. This is consistent with the evolution of stress in the strain-rate jump tests carried out at 400ºC for the 7, 16 and 34 nm thick multilayers, that clearly reach a "steady-state" constant stress $\sigma_{ss}$ after each strain-rate jump. This is contrary to the room temperature behavior and the behaviour of the 63 nm thick multilayer at both temperatures, which show strain-hardening during the entire compression tests.

Under steady-state conditions, the strain rate $\dot{\varepsilon}$ of polycrystalline materials depends on the steady-state stress $\sigma_{ss}$ and temperature $T$ according to

$$\dot{\varepsilon} = A \frac{\sigma_{ss}^n}{d^p} \exp\left(\frac{-Q_c}{RT}\right) \qquad (7)$$

where $A$ is a constant, $d$ the grain size, $n$ the stress exponent ($n$ = 1 for Coble creep, $n$ = 2 for grain boundary sliding and $n$ = 3-7 for dislocation creep), $p$ the grain boundary exponent ($p$ = 3 for Coble creep, $p$ = 2 for grain boundary sliding and $p$ = 0 for dislocation creep), and $Q_c$ the activation energy for creep. The strain rate versus the steady-state stress is plotted in Fig. 10a for the 7, 16 and 34 nm thick multilayers, that seem to be controlled by stress-assisted diffusion processes. All the data fit well with a stress exponent $n \approx 7$. This value is indicative of dislocation climb power-law creep, even though the temperature is relative low ($\approx 0.3\ T_m$). Moreover, if the layer thickness, $h$, is taken as the characteristic length scale for NMMs, instead of the grain size, $d$, the steady-state stress at constant strain rate varies with $h^{0.65}$, as shown in Fig. 10b, that plots the steady-state stress versus layer thickness at constant strain rate. This layer thickness dependency indicates an increasing diffusion activity for thinner layer thicknesses, in agreement with the higher density of Cu globules along the side length of the pillars for the thinnest layers shown in Fig. 5. So, it is speculated that dislocation climb at interfaces, controlled by grain boundary diffusion, might be the rate controlling deformation mechanism at 400ºC for the nanolaminates with layer thicknesses < 34 nm. This is in agreement with atomistic modelling studies by Wang et al. [1], which predicted dislocation climb at interfaces in PVD Cu/Nb and experimental observations by *in-situ* indentation on Al-Nb multilayers in a TEM [34].

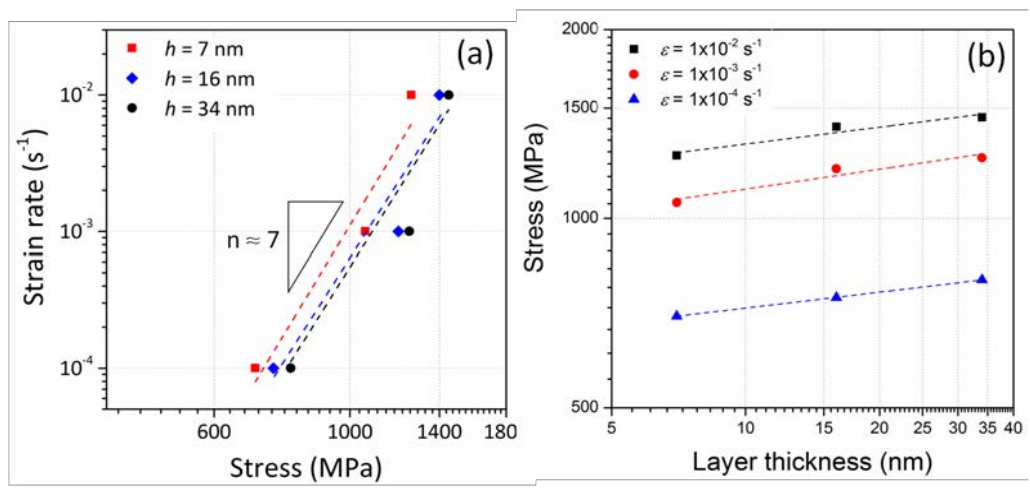

**Fig. 10.** (a) Log-log plot of strain rate versus stress for layer thicknesses between 7 and 34 nm at 400ºC. The experimental data are in agreement with a stress exponent $n \approx 7$. (b) Log-log plot of steady-state stress at constant strain rate versus layer thickness. The experimental data are in agreement with a grain size exponent $p \approx 0.65$.

4.3. Deformation map of ARB Cu/Nb.

Based on these results, a deformation mechanism map for ARB Cu/Nb multilayers is proposed in Fig. 11 as a function of layer thickness at 25ºC and 400ºC for strain rates in the range $10^{-2}$-$10^{-4}$ s$^{-1}$. The blue and red lines represent the flow stress evolution with layer thickness at 25ºC and 400ºC, respectively. Three mechanisms are considered:

- Confined layer slip, CLS, controls deformation for h ≳ 20 nm at 25ºC and for h ≳ 50 nm at 400ºC, following equation (6). Under CLS, the strength reduction with temperature is limited and the strain rate sensitivity is small: $m \approx 0.006$ at room temperature and $m \approx 0.05$ at 400ºC, as expected in a glide controlled regime. No purely steady-state conditions are reached at the strain rates considered at room temperature in Fig. 4, but the strain hardening rate is almost negligible for strains > 5%, so the flow stresses at the plateau for each strain rate are plotted in Fig. 11.
- Dislocation transmission at interfaces, DT, controls deformation for h < 20 nm at 25ºC. The stress required for dislocation transmission was estimated to be ≈ 1.7 GPa from the yield stress of the 7 nm nanolamiante, but the flow stresses reaches quickly 2.3 GPa for a strain rate of $10^{-3}$ s$^{-1}$ for applied strains > 5%. The flow stresses at the plateau are thus plotted in Fig. 11, considering a slightly larger strain rate sensitivity, $m \approx 0.010$, than for CLS.
- Dislocation climb controlled by interface diffusion, DC, controls deformation for h < 25-35 nm at 400ºC. From equation (7), the steady-state flow stress at 400ºC can be expressed as:

$$\sigma_{\dot{\varepsilon}} = [K \cdot \dot{\varepsilon} \cdot h^{0.65} \cdot ]^{1/7} \qquad (8)$$

where $K$ at 400ºC is determined from the experimental results but should be related to the activation energy for creep, $Q_c$. $Q_c$ could not be determined due to the lack of more elevated temperature data, but it should be close to the activation energy for grain boundary diffusion in Cu [33], ~62-104 kJ/mol based on the experimental observations.

The deformation map addresses that the flow stresses of ARB Cu/Nb NMMs with layer thicknesses h > 25-35 nm at 400ºC is limited by stress-assisted diffusion. The exact layer thickness for which the transition from CLS to DC takes place depends on the applied strain rate. As a result, this regime is characterized by a "smaller is softer" inverse size effect.

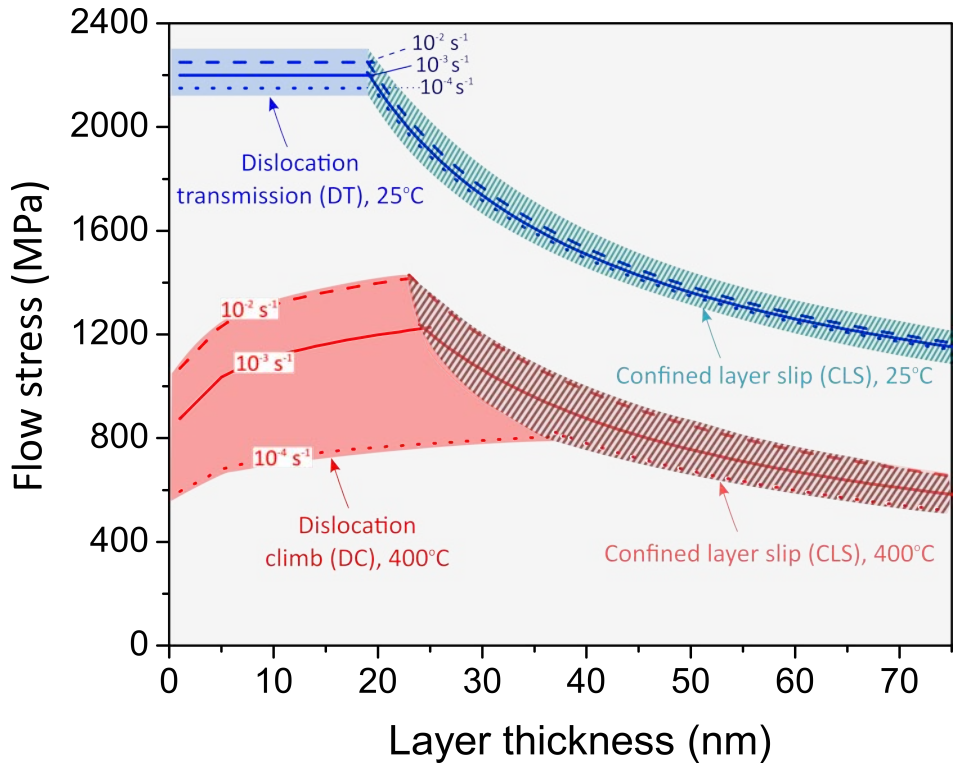

**Fig. 11.** Strain room-temperature and high-temperature deformation mechanism map for ARB Cu/Nb NMMs with layer thickness below 75 nm. Map is valid for a strain-rates in the range $10^{-2}$ -$10^{-4}$ s$^{-1}$.

## 5. Conclusions

Micropillar compression tests at room temperature of ARB Cu/Nb multilayers with layer thicknesses (*h*) of 7, 16, 34 and 63 nm revealed an increase in strength as the layer thickness decreased. Deformation was controlled by confined layer slip in the nanolaminates with thicker layers and by dislocation transmission at interfaces in the thinner ones. Similar tests at 400 °C revealed a reduction in strength which increased as the layer thickness decreased. The strength drop was particularly large for layer thicknesses of 7 nm and 16 nm and stress-assisted diffusion along interfaces was found to play a dominant role in these nanolaminates. The enhanced strain-rate sensitivities, with a stress exponent of n ≈ 7 and the low activation volumes indicated diffusion-assisted dislocation climb to be the rate-controlling mechanism. For increasing layer thickness and higher strain-rates, diffusion-related processes were found to decrease suggesting a transition to deformation controlled dislocation glide.

A deformation mechanism map is proposed for ARB Cu/Nb NMMs as a function of layer thickness for 25ºC and 400ºC for strain ranges in the range $10^{-2}$-$10^{-4}$ s$^{-1}$, accounting for confined layer slip (CLS), dislocation transmission (DT) at interfaces and dislocation climb (DC). The map shows a transition from CLS to DT at room temperature for layer thicknesses below 20 nm and a transition from CLS to DC at 400ºC for layer thicknesses below 25-35 nm, depending on applied strain rate.

The limited strength drop at 400ºC for nanolaminates with larger layer thickness indicates an optimum elevated temperature strength for layer thicknesses in the range 25-35 nm. Below this critical thickness, the additional strength gained by a layer thickness reduction at room temperature is exceeded by the reduction in strength due to the activation of stress-assisted diffusion processes at 400ºC.

**Acknowledgements**

This investigation was supported by the European Research Council (ERC) under the European Union's Horizon 2020 research and innovation programme (Advanced Grant VIRMETAL, grant agreement No. 669141). IJB acknowledges financial support from the National Science Foundation (NSF CMMI-1729887). Useful discussions during the course of this work from Prof. Sybrand van der Zwaag are gratefully acknowledged. This work was performed, in part, at the Center for Integrated Nanotechnologies, an Office of Science User Facility operated for the U.S. Department of Energy (DOE) Office of Science. Los Alamos National Laboratory, an affirmative action equal opportunity employer, is operated by Los Alamos National Security, LLC, for the National Nuclear Security Administration of the U.S. Department of Energy under contract DE-AC52-06NA25396.

# References

[1] J. Wang, A. Misra, *Curr. Opin. Solid State Mater. Sci.* 15, 20 (2011).
[2] A. Misra, J.P. Hirth, R.G. Hoagland, *Acta Mater.* 53, 4817 (2005).
[3] N.A. Mara, D. Bhattacharyya, R.G. Hoagland, A. Misra, *Scr. Mater.* 58, 874 (2008).
[4] K. Hattar, A. Misra, M.R.F. Dosanjh, P. Dickerson, I.M. Robertson, R.G. Hoagland, *J. Eng. Mater. Technol.* 134, 21014 (2012).
[5] J.S. Carpenter, S.J. Zheng, R.F. Zhang, S.C. Vogel, I.J. Beyerlein, N.A. Mara, *Philos. Mag.* 93, 718 (2013).
[6] S. Zheng, I.J. Beyerlein, J.S. Carpenter, K. Kang, J. Wang, W. Han, N.A. Mara, *Nat. Commun.* 4, 1696 (2013).
[7] A. Misra, M.J. Demkowicz, X. Zhang, R.G. Hoagland, *JOM.* 59, 62 (2007).
[8] N.A. Mara, A. Misra, R.G. Hoagland, A.V. Sergueeva, T. Tamayo, P. Dickerson, A.K. Mukherjee, *Mater. Sci. Eng. A* 493, 274 (2008).
[9] N.A. Mara, A. Sergueeva, A. Misra, A.K. Mukherjee, *Scr. Mater.* 50, 803 (2004).
[10] N.A. Mara, T. Tamayo, A.V. Sergueeva, X. Zhang, A. Misra, A.K. Mukherjee, *Thin Solid Films* 515, 3241 (2007).
[11] A. Misra, R.G. Hoagland, H. Kung, *Philos. Mag.* 84, 1021 (2004).
[12] A. Misra, R.G. Hoagland, *J. Mater. Res.* 20, 2046 (2011).
[13] S.J. Zheng, J. Wang, J.S. Carpenter, W.M. Mook, P.O. Dickerson, N.A. Mara, I.J. Beyerlein, *Acta Mater.* 79, 282 (2014).
[14] M.A. Monclús, S.J. Zheng, J.R. Mayeur, I.J. Beyerlein, N.A. Mara, T. Polcar, J. Llorca, J.M. Molina-Aldareguía, *APL Mater.* 1, 52103 (2013).
[15] J.S. Carpenter, S.C. Vogel, J.E. LeDonne, D.L. Hammon, I.J. Beyerlein, N.A. Mara, *Acta Mater.* 60, 1576 (2012).
[16] G. Mohanty, J.M. Wheeler, R. Raghavan, J. Wehrs, M. Hasegawa, S. Mischler, L. Philippe, J. Michler, *Philos. Mag.* 95, 1878 (2014).
[17] I.N. Sneddon, *Int. J. Eng. Sci.* 3, 47 (1965).
[18] S.J. Zheng, I.J. Beyerlein, J. Wang, J.S. Carpenter, W.Z. Han, N.A. Mara, *Acta Mater.* 60 5858, (2012).
[19] S. Lotfian, M. Rodríguez, K.E. Yazzie, N. Chawla, J. Llorca, J.M. Molina-Aldareguía, *Acta Mater.* 61, 4439 (2013).
[20] R. Raghavan, J.M. Wheeler, D. Esqué-de los Ojos, K. Thomas, E. Almandoz, G.G. Fuentes, J. Michler, *Mater. Sci. Eng. A* 620, 375 (2014).
[21] R. Raghavan, J.M. Wheeler, T.P. Harzer, V. Chawla, S. Djaziri, K. Thomas, B. Philippi, C. Kirchlechner, B.N. Jaya, J. Wehrs, J. Michler, G. Dehm, *Acta Mater.* 100, 73 (2015).
[22] X.Y. Zhu, J.T. Luo, F. Zeng, F. Pan, *Thin Solid Films* 520, 818 (2011).
[23] A. Misra, R.G. Hoagland, *J. Mater. Sci.* 42, 1765 (2007).
[24] Y.A. Chang, L. Himmel, *J. Appl. Phys.* 37, 3567 (1966).
[25] R.L. Coble, *J. Appl. Phys.* 34, 1679 (1963).
[26] O.A. Ruano, O.D. Sherby, J. Wadsworth, J. Wolfenstine, *Mater. Sci. Eng. A* 211, 66 (1996).
[27] T. Surholt, C. Herzig, *Acta Mater.* 45, 3817 (1997).
[28] D.B. Butrymowicz, J.R. Manning, M.E. Read, *J. Phys. Chem. Ref. Data 2*, 643 (1973).
[29] M.A. Meyers, A. Mishra, D.J. Benson, A. Mishra, *Prog. Mater. Sci.* 51, 427 (2006).
[30] Q. Wei, S. Cheng, K.T. Ramesh, E. Ma, *Mater. Sci. Eng. A* 381, 71 (2004).
[31] M.E. Kassner, M.T. Pérez-Prado, *Prog. Mater. Sci.* 45, 1 (2000).
[32] J. Wehrs, G. Mohanty, G. Guillonneau, A.A. Taylor, X. Maeder, D. Frey, L. Philippe, S. Mischler, J.M. Wheeler, J. Michler, *JOM.* 67, 1684 (2015).
[33] H.J. Frost, M.F. Ashby, *Deformation-Mechanism Maps : The Plasticity and Creep of Metals and Ceramics*, (Pergamon Press, Oxford; New York, 1982).